\documentclass[conference]{IEEEtran}
\ifCLASSINFOpdf
\else
\fi
\usepackage{url}
\usepackage{amsbsy}
\usepackage{amsmath}
\usepackage{fixmath}
\usepackage{amssymb}
\usepackage{epsfig}
\usepackage{epstopdf}
\usepackage{algorithmic}
\usepackage{algorithm}

\usepackage{tikz}
\usetikzlibrary{arrows,decorations,backgrounds,shadows,plotmarks,calc}

\usepackage{pgfplots}
\pgfplotsset{compat=newest}
\pgfplotsset{plot coordinates/math parser=false}
\pgfplotsset{every axis/.append style={font=\footnotesize}}
\pgfplotsset{
    ylabel right/.style={
        after end axis/.append code={
            \node [rotate=90, anchor=north] at (rel axis cs:1,0.5) {#1};
        }   
    }
}
\newlength\figureheight
\newlength\figurewidth
\newlength\subgraphheight
\newlength\subgraphwidth
\makeatletter
\pgfdeclareshape{dspmixer}{
\inheritsavedanchors[from=circle]
    \inheritanchorborder[from=circle]

    \inheritanchor[from=circle]{center}
    \inheritanchor[from=circle]{south}
    \inheritanchor[from=circle]{west}
    \inheritanchor[from=circle]{north}
    \inheritanchor[from=circle]{east}
    \inheritanchor[from=circle]{south east}
    \inheritanchor[from=circle]{south west}
    \inheritanchor[from=circle]{north east}
    \inheritanchor[from=circle]{north west}

    \inheritbackgroundpath[from=circle]

    \beforebackgroundpath{

        \radius \pgf@xb=
        0.7071067812\pgf@x
        \centerpoint \pgf@xa=\pgf@x \pgf@ya=\pgf@y \pgf@xc=\pgf@x \pgf@yc=\pgf@y

        \advance\pgf@xa by -\pgf@xb
        \advance\pgf@ya by -\pgf@xb
        \advance\pgf@xc by \pgf@xb
        \advance\pgf@yc by \pgf@xb
        \pgfpathmoveto{\pgfpoint{\pgf@xa}{\pgf@ya}}
        \pgfpathlineto{\pgfpoint{\pgf@xc}{\pgf@yc}}

        \advance\pgf@ya by 2\pgf@xb
        \advance\pgf@yc by -2\pgf@xb
        \pgfpathmoveto{\pgfpoint{\pgf@xa}{\pgf@ya}}
        \pgfpathlineto{\pgfpoint{\pgf@xc}{\pgf@yc}}

        \pgfusepath{draw}
    }
}
\makeatother
\DeclareMathOperator*{\h}{H}

\DeclareMathOperator*{\E}{E}

\begin{document}
\title{Compensation of Amplifier Distortion for OFDM signals based on Iterative Hard Thresholding}

\author{\IEEEauthorblockN{Wasim Amjad, Javier Garc\'ia, Jawad Munir, Amine Mezghani and Josef A. Nossek}
	\IEEEauthorblockA{Institute for Circuit Theory and Signal Processing\\
		Munich University of Technology, 80290 Munich, Germany\\
		E-Mail: \{wasim.bharah, javier.garcia, jawad.munir, amine.mezghani, josef.a.nossek\}@tum.de}
%\and
%\IEEEauthorblockN{Homer Simpson}
%\IEEEauthorblockA{Twentieth Century Fox\\
%\and
%\IEEEauthorblockN{James Kirk\\ and Montgomery Scott}
%\IEEEauthorblockA{Starfleet Academy\\
%San Francisco, California 96678-2391\\
%Fax: (888) 555--1212}
}


%Telephone: (800) 555--1212, Fax: (888) 555--1212}


\maketitle

\begin{abstract}
 The mitigation of nonlinear distortion caused by power amplifiers (PA) in Orthogonal Frequency Division Multiplexing
(OFDM) systems is an essential issue to enable energy efficient operation. In this work we proposed a new algorithm for receiver-based clipping estimation in OFDM systems that combines the Iterative Hard Thresholding method introduced in~\cite{blumensath2009iterative} with the weighting corresponding to the estimated probability of clipping used in~\cite{Ali2014wpasamp}. Thereby a more general amplifier input-output characteristic is considered, which is assumed to be unknown at the receiver side. Further, we avoid the use of dedicated subcarriers and formulate the recovery problem solely on reliably detected sub-carriers. Through simulations, we show that the proposed technique achieves a better complexity-performance tradeoff compared to existing methods.
 
\end{abstract}


% \fi
\IEEEpeerreviewmaketitle

\section{Introduction}\label{sec:intro}

With the recent leap from 3G to 4G, the amount of power consumption in mobile devices has increased considerably. The
4G operates on Long Term Evolution (LTE) in which Orthogonal Frequency Division Multiplexing
(OFDM) is used for the wireless communications. However, the use of multi-carrier
modulation such as OFDM leads to the higher Peak-to-Average Power (PAPR) and thus
higher power consumption in an RF Power Amplifier (PA) as well as the Digital-to-Analog
(ADC) converters in the mobile devices. In a typical 4G system, RF PAs have to satisfy stringent linearity requirements and thus accounts for a significant amount of the total power consumption. 
Therefore, in order to  operate the amplifiers at maximal possible efficiency and reduce the requirements on the DAC, it is important  to allow a partial clipping of the OFDM signal, while envisaging at the same time the compensation of the resulting distortion at the receiver side. In fact, when operating at such lower input back-off for the PA results also into a certain out-off-band radiation and a certain in-band distortion. This paper consider the compensation of the in-band distortion.

In this context, the work \cite{Gregorio2012} investigated the iterative maximum likelihood (ML) detection of the clipped OFDM signal based on the turbo decoding approach and developed an computationally efficient algorithm called power amplifier nonlinearity cancellation (PANC). The main disadvantage of this method is that the  input-output characteristic of the PA has to be perfectly known at the receiver side.

In anther recent line of work, a sparse
reconstruction algorithm, called Support Agnostic Bayesian Matching Pursuit (SABMP) has been proposed in \cite{Al-Safadi2012,al2013receiver,6292976,Ali2014wpasamp} for the recovery of clipped OFDM signals has been investigated based on the sparse nature of the clipping error while choosing a certain number reliable sub-carrier  for Setting up the sparse recovery problem.  Nevertheless, it has been observed, that this algorithm has relatively
higher complexity. Therefore, a new sparse reconstruction approach based on Weighted Iterative Hard Thresholding (WIHT) is proposed 
which has significant lower computation complexity than that of SABMP for almost the
same performance. Moreover, a more general amplifier characteristic is introduced and analyzed, which is assumed to unknown at the receiver. Finally, the proposed sparse
reconstruction algorithm is applied and simulated for a single-input single-output (SISO) system.

Our paper is organized as follows. Section~\ref{sec:channel} describes the general system model including the amplifier model and formulates the recovery problem. In Section~\ref{sec:wiht}, the proposed  Weighted Iterative Hard Thresholding algorithm is presented. Section~\ref{sec:complexity} compares the computational complexity of this technique with two other widely used methods, namely Support Agnostic Bayesian Matching Pursuit (SABMP)~\cite{Ali2014wpasamp} and Power Amplifier Nonlinearity Cancellation (PANC)~\cite{Gregorio2012}. In Section~\ref{sec:results} we provide some simulation results to reveal the usefulness of the method, and Section~\ref{sec:conclusion} gives a brief conclusion and suggests some areas for future work.
 
\section{System Model}\label{sec:channel}
Let us consider an OFDM system in which the incoming information bits are mapped onto $L$-ary QAM constellation and concatenated to form an $N$-dimensional frequency domain data symbols vector $\boldsymbol{\mathcal{X}}\in \mathbb{C}^{N\times1}$. The time-domain vector $\mathbf{x}$ can be obtained by $\mathbf{x} = \mathbf{F}^H\boldsymbol{\mathcal{X}}$, where $\mathbf{F}$ is a $N\times N$ DFT matrix \cite{al2013receiver}.

The time-domain signal $\mathbf{x}$ has high PAPR due to the addition of different frequency sub-carriers. Because of this, the power amplifier (PA) will operate in the non-linear region, introducing clipping for the higher values of $\mathbf{x}$. The clipping model we are considering is a generalization of the soft clipping used in~\cite{Al-Safadi2012} in that it also introduces phase distortion in the non-linear region. We motivate this by the fact that the behavior in this region is generally unpredictable and cannot always be compensated, while phase distortion in the linear region is usually mild and can be compensated with pre-distortion. The samples of the clipped signal $\mathbf{x_p}$ are then given by:
\begin{equation}
	x_p(i)=\begin{cases}
		x(i),&\left|x(i)\right|\leq\tau\\
		\tau e^{j \left( \arg{x(i)}+\phi\left(\left|x(i)\right|\right)\right)}  &  \left|x(t)\right|>\tau
	\end{cases}
	\label{eq:pa_model_nopreserve}
\end{equation}
where $\tau$ is the clipping threshold and $\phi\left(\left|x(i)\right|\right)$ is a phase distortion function defined as the argument of the Band 1 PA model given by 3GPP in~\cite{3gpp_pa_model}. We note that our proposed sparse recovery technique is able to recover the clipping without any prior knowledge of this phase distortion function.

The clipping can be seen as the addition of a sparse signal $\mathbf{c}$ in the time-domain signal $\mathbf{x}$ and can be written as

\begin{equation}
	\mathbf{x_p} = \mathbf{x} + \mathbf{c},
\end{equation}
which can be written as follows
\begin{equation}
	\mathbf{x_p} = \mathbf{F^H}\boldsymbol{\mathcal{X}} + \mathbf{c}.
\end{equation}	

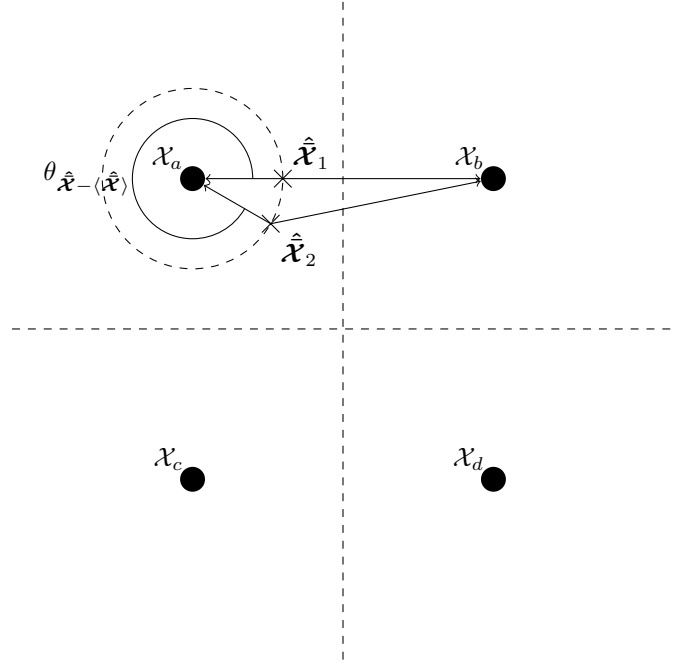
\begin{figure}[!t]\centering
\begin{tikzpicture}[scale=2,
ypoint/.style={
	dspmixer,
	minimum width=0.05
},
qpoint/.style={
	circle,
	minimum width=0.05,
	fill
}
]

% constellation points
\node (qn1n1) [qpoint] at (-1, -1) {};
\node (qn11) [qpoint] at (-1, 1) {};
\node (q11) [qpoint] at (1, 1) {};
\node (q1n1) [qpoint] at (1, -1) {};
% RX symbol points
\node (y1) [ypoint] at ($(qn11)+(0:0.6)$) {};
\node (y2) [ypoint] at ($(qn11)+(-30:0.6)$) {};

% constellation labels
\node [anchor=south east] at (qn11) {$\mathcal{X}_a$};
\node [anchor=south east] at (q11) {$\mathcal{X}_b$};
\node [anchor=south east] at (qn1n1) {$\mathcal{X}_c$};
\node [anchor=south east] at (q1n1) {$\mathcal{X}_d$};
% RX symbol labels
\node [anchor=south west] at (y1) {$\boldsymbol{\hat{\bar{\mathcal{X}}}}_1$};
\node [anchor=north west] at (y2) {$\boldsymbol{\hat{\bar{\mathcal{X}}}}_2$};

% constellation regions
\draw[dashed] (0, -2.2) -- (0, 2.2);
\draw[dashed] (-2.2, 0) -- (2.2, 0);
% same distance cirunference
\draw[dashed] (qn11) circle (0.6);

% distance arrows
\path[arrows={->}]
	(y1.center) edge (qn11)
				edge (q11)
	(y2.center) edge (qn11)
				edge (q11);
				
% theta arc
\draw ($(qn11)+(0.4, 0)$) arc (0:330:0.4);
\node [anchor=east] at ($(qn11)+(-0.35, 0)$) {$\theta_{\boldsymbol{\hat{\bar{\mathcal{X}}}} - \langle \boldsymbol{\hat{\bar{\mathcal{X}}}}\rangle}$};	
\end{tikzpicture}
			\caption{Reliability of observation relevant to distance and angle to different constellation points}
			\label{fig:reliable_carriers}			
		\end{figure}

In order to avoid the ISI at the receiver, a cyclic prefix (CP) is appended to the time-domain signal. This CP is removed at the receiver and then FFT is performed at the receiver to obtain the received signal as
\begin{equation}\label{y}
	\mathbf{y} = \mathbf{Hx_p} + \mathbf{z},
\end{equation}
where $\mathbf{z}$ is the AWGN noise with the variance $\sigma_n^2$. $\mathbf{H}$ is the circulant channel matrix due the CP insertion/removal property and is given as $\mathbf{H} = \mathbf{F^H\Lambda F}$. The received time-domain signal $\mathbf{y}$ is then written as follows

\begin{equation}
\mathbf{y} = \mathbf{F^H\Lambda Fx_p+z}.
\end{equation}
The received signal in the frequency domain can then be written as
\begin{equation}\label{obstacle}
\boldsymbol{\mathcal{Y}} = \mathbf{\Lambda (\boldsymbol{\mathcal{X}}+\boldsymbol{\mathcal{C}})}+\boldsymbol{\mathcal{Z}},
\end{equation}
the estimated received signal is obtained by equalizing the channel as follows
\begin{equation}
\boldsymbol{\hat{\bar{\mathcal{X}}}} = \mathbf{\Lambda^{-1}}\boldsymbol{\mathcal{Y}} = \mathbf{ (\boldsymbol{\mathcal{X}}+\boldsymbol{\mathcal{C}})}+\mathbf{\Lambda^{-1}}\boldsymbol{\mathcal{Z}},
\end{equation}
In order to determine the unknown vector $\mathbf{c}$, $\boldsymbol{\mathcal{X}}$ needs to be eliminated. The process for elimination is carried out by projecting onto a set of reliable carriers. This is the set of carriers where we have more certainty that the distortion introduced by noise and clipping was not big enough to move the symbol to a different decision region in the constellation. The reliable carriers are chosen according to a reliability measure that depends on both the distance to the closest constellation point and the angle to it: 

\begin{multline}
\mathcal{R}(\boldsymbol{\hat{\bar{\mathcal{X}}}} - \langle \boldsymbol{\hat{\bar{\mathcal{X}}}}\rangle) = \frac{\sqrt{2}d_{\text{min}}-|\boldsymbol{\hat{\bar{\mathcal{X}}}} - \langle \boldsymbol{\hat{\bar{\mathcal{X}}}}\rangle|}{\sqrt{2}d_{\text{min}}} + \\ 
+ \frac{|\boldsymbol{\hat{\bar{\mathcal{X}}}} - \langle \boldsymbol{\hat{\bar{\mathcal{X}}}}\rangle|}{\sqrt{2}d_{\text{min}}}\cos(4\theta_{\boldsymbol{\hat{\bar{\mathcal{X}}}} - \langle \boldsymbol{\hat{\bar{\mathcal{X}}}}\rangle}+\pi),
\end{multline} 

where $d_{min}$ is the distance between two constellation neighbors, $\langle \boldsymbol{\hat{\bar{\mathcal{X}}}}\rangle$ is the closest constellation point to $\boldsymbol{\hat{\bar{\mathcal{X}}}}$, and $\theta_{\boldsymbol{\hat{\bar{\mathcal{X}}}} - \langle \boldsymbol{\hat{\bar{\mathcal{X}}}}\rangle}$ is the angle to it, as depicted in Fig.~\ref{fig:reliable_carriers}.  This geometric measure was taken from~\cite{6292976}, and is motivated by the fact that a deviation from the constellation point in an angle close to $k\frac{\pi}{2}, k\in\{0, 1, 2, 3\}$ makes the subcarrier less reliable than a deviation in an angle close to $(2k+1)\frac{\pi}{4}, k\in\{0, 1, 2, 3\}$, due to the fact that the distance to the second closest constellation point is higher in the second case. In summary, samples that have a shorter distance to their closest constellation point, or whose angle to it is closer to $(2k+1)\frac{\pi}{4}, k\in\{0, 1, 2, 3\}$ will get a higher reliability value. The $P$ subcarriers with highest value of $\mathcal{R}(\boldsymbol{\hat{\bar{\mathcal{X}}}} - \langle \boldsymbol{\hat{\bar{\mathcal{X}}}}\rangle)$ are chosen as reliable.

Taking the difference between the estimate $\boldsymbol{\hat{\bar{\mathcal{X}}}}$ and the associated vector $\langle \boldsymbol{\hat{\bar{\mathcal{X}}}}\rangle$ after the slicer and then projecting onto the chosen reliable carriers yields

\begin{align}
\boldsymbol{J}(\boldsymbol{\hat{\bar{\mathcal{X}}}}-\langle \boldsymbol{\hat{\bar{\mathcal{X}}}}\rangle) &=& &\boldsymbol{J}(\mathbf{ (\boldsymbol{\mathcal{X}}+\boldsymbol{\mathcal{C}}-\langle \boldsymbol{\hat{\bar{\mathcal{X}}}}\rangle)})+\boldsymbol{J}\mathbf{\Lambda^{-1}}\boldsymbol{\mathcal{Z}} \\
&=& &\boldsymbol{J}\boldsymbol{\mathcal{C}}+\boldsymbol{J}\mathbf{\Lambda^{-1}}\boldsymbol{\mathcal{Z}}\\
&=& &\boldsymbol{J}\mathbf{Fc}+\boldsymbol{J}\mathbf{\Lambda^{-1}}\boldsymbol{\mathcal{Z}},
\end{align}
where $\boldsymbol{J}$ is an $P\times N$ binary selection matrix indicating the locations of the reliable carriers. Above equation can then be written as

\begin{equation}\label{eq: final_model}
\bar{\boldsymbol{\mathcal{Y}}} = \mathbf{Ac} + \boldsymbol{\mathcal{Z}}',
\end{equation}
where $\bar{\boldsymbol{\mathcal{Y}}}=\boldsymbol{J}(\boldsymbol{\hat{\bar{\mathcal{X}}}}-\langle \boldsymbol{\hat{\bar{\mathcal{X}}}}\rangle)$, $\mathbf{A} = \boldsymbol{J}\mathbf{F}$ and $\boldsymbol{\mathcal{Z}}' = \boldsymbol{J}\mathbf{\Lambda^{-1}}\boldsymbol{\mathcal{Z}}$. 

This is a compressed sensing model that can now be processed for the unknown $\mathbf{c}$ by using any CS algorithms. In the following, an efficient CS algorithm will be used for determining the unknown sparse vector $\mathbf{c}$. Note that, in contrast to~\cite{al2013receiver}, we do not include the phase information into the sensing matrix $\mathbf{A}$ and recover a complex sparse vector $\mathbf{c}$ in order to be able to compensate also for phase distortion in the non-linear region of the PA.

\section{Weighted Iterative Hard Thresholding (WIHT)} \label{sec:wiht}
The IHT algorithm~\cite{blumensath2009iterative} is an efficient greedy technique for the sparse recovery problem. However, it cannot be directly applied for the CS model in (\ref{eq: final_model}) and some modifications are required in order to make use of this robust method. We call our proposed modified algorithm Weighted Iterative Hard Thresholding (WIHT).

The summary of the proposed technique is shown in Algorithm \ref{algo:WIHT}. The algorithms takes  $\boldsymbol{\hat{\bar{\mathcal{X}}}}$,  $\mathbf{\boldsymbol{\bar{\mathcal{Y}}}}$, $\mathbf{A}$ and the estimated number $K$ of active taps in $\mathbf{c}$  as inputs. We note that the performance of our method does not degrade noticeably if this value is overestimated, therefore, as proposed in~\cite{Masood2012}, we choose $K$ to be slightly higher than the number of elements of $\left|\mathbf{A}^H\boldsymbol{\overline{\mathcal{Y}}}\right|$ that are greater or equal than half its maximum value.

\begin{equation}
K=\left|\left\{j:\left|\mathbf{a}_j^H\boldsymbol{\overline{\mathcal{Y}}}\right|\geq\frac{1}{2}\left\|\mathbf{A}^H\boldsymbol{\overline{\mathcal{Y}}}\right\|_{\infty}\right\}\right|,
\end{equation}

where $\mathbf{a}_j$ denotes the $j$-th column of $\mathbf{A}$. The clipping threshold $\tau$ is approximated by taking the maximum absolute value of the estimated clipped signal $\mathbf{\hat{x}_p}=\mathbf{F}^{\mathrm{H}}\boldsymbol{\hat{\bar{\mathcal{X}}}}$. Afterwards, a weighting vector $\mathbf{w}$ is calculated to approximate the probability that each sample was clipped. As this probability must be higher for samples closer to the clipping threshold, and be constrained to the range $\left[0, 1\right]$, a negative exponential of the difference between the approximated threshold  $\hat{\tau}$ and the estimated clipped signal value was chosen as the weighting function~\cite{Ali2014wpasamp}:

\begin{equation}
\mathbf{w} = \exp{\left\{-\left({\hat{\tau}} - |\mathbf{\hat{x}_p}|\right)\right\}}.
\end{equation}

 $\mathbf{W}$ is the diagonal matrix with the weighting entries along its diagonal. The function $H_K(\cdot)$ as shown in the algorithm below can be defined as
 
 \begin{equation}
 H_K(\mathbf{A},\mathbf{b}) =\begin{cases}
 \mathbf{A}^H\mathbf{b},&\mathbf{A}^H\mathbf{b} \geq \text{max}_K(\mathbf{A}^H\mathbf{b})\\
 0  &  \text{otherwise},
 \end{cases}
 \label{eq:Hk}
 \end{equation}
 where $\text{max}_k(\cdot)$ gives the $k$-th maximum value. In other words, the function $H_K(\mathbf{A},\mathbf{b})$ is used to find out the $k$ maximum values of the correlation between the matrices $\mathbf{A}$ and the vector $\mathbf{b}$. The remaining correlation values which are not included in the $k$ maximum values are then forced to zero. In the algorithm below, the correlation is computed between the measurement matrices $\mathbf{A}$ and the observation vector ${\bar{\mathcal{Y}}}$ and then the support of the sparse vector is computed by simply taking the support of the output of the function $H_k(\cdot)$.
 
 Up until now, a single iteration of standard IHT has been run. We found out that the values of the taps of the sparse vector $\mathbf{c}$ are generally not well approximated after only one iteration, but the support of $\boldsymbol{\hat{\bar{c}}}^{\left(1\right)}$ usually coincides with that of $\mathbf{c}$. Therefore, Weighted IHT takes the support $\mathcal{S}_{\text{IHT}}$ of $\boldsymbol{\hat{\bar{c}}}^{\left(1\right)}$ and performs a Best Linear Unbiased Estimate (BLUE) of $\mathbf{c}$ over it: 
 \begin{equation}
 \mathbf{\hat{\bar{c}}}=(\mathbf{A_{\mathcal{S}_{\text{IHT}}}^{\mathrm{H}}A_{\mathcal{S}_{\text{IHT}}}})^{-1}\mathbf{A_{\mathcal{S}_{\text{IHT}}}^{\mathrm{H}}}\boldsymbol{\bar{\mathcal{Y}}},
 \end{equation}
 where $\mathbf{A}_{\mathcal{S}_{\text{IHT}}}=\boldsymbol{J_{\mathcal{S}_{\text{IHT}}}}\mathbf{A}$ is obtained by taking the columns of $\mathbf{A}$ whose indices are in the support $\mathcal{S}_{\text{IHT}}$. This BLUE estimate approximates the conditional expectation over the observation and the computed support, $\mathbf{\hat{\bar{c}}}\approx\mathbb{E}\left[\mathbf{c}|\boldsymbol{\bar{\mathcal{Y}}},\mathcal{S}_{\text{IHT}}\right]$. Finally, the actual pre-clipped transmitted signal can be obtained by subtracting the estimated sparse vector $\mathbf{\hat{\bar{c}}}$ from the estimated clipped signal $\mathbf{\hat{x}_p}$.
\begin{algorithm}
	\caption{Weighted Iterative Hard Thresholding (WIHT)}
	\label{algo:WIHT}
	\begin{algorithmic}
		\REQUIRE $\boldsymbol{\hat{\bar{\mathcal{X}}}}, \mathbf{\boldsymbol{\bar{\mathcal{Y}}}}$, $\mathbf{A}$, $K$
		\STATE \textbf{Estimate of clipped signal:} \quad $\mathbf{\hat{x}_p}=\mathbf{F}^{\mathrm{H}}\boldsymbol{\hat{\bar{\mathcal{X}}}}$
		\STATE \textbf{Threshold estimation:} \quad  $\hat{\tau} = \|\mathbf{\hat{x}_p}\|_\infty$
		\STATE \textbf{Weighting vector:} \quad $\mathbf{w} = \exp{\left\{-\left({\hat{\tau}} - |\mathbf{\hat{x}_p}|\right)\right\}}$
		\STATE \textbf{Weighting matrix: }
		$\mathbf{W} = \text{diag}\{\mathbf{w}\}$
		\STATE \textbf{Support of Sparse Vector: } $\mathcal{S}_{\text{IHT}} = \text{Supp}(H_K\left(\mathbf{W}\mathbf{A}^{\h}\mathbf{\boldsymbol{\bar{\mathcal{Y}}}}\right))$
		\STATE \textbf{BLUE Estimate: }
		$\boldsymbol{\hat{\bar{c}}} = \mathbb{E}[\mathbf{c}|\boldsymbol{\bar{\mathcal{Y}}},\mathcal{S}_{\text{IHT}}] \approx (\mathbf{A_{\mathcal{S}_{\text{IHT}}}^{\mathrm{H}}A_{\mathcal{S}_{\text{IHT}}}})^{-1}\mathbf{A_{\mathcal{S}_{\text{IHT}}}^{\mathrm{H}}}\boldsymbol{\bar{\mathcal{Y}}}$
		\ENSURE $\mathbf{\hat{x}} = \mathbf{\hat{x}_p} - \mathbf{\hat{\bar{c}}}$
	\end{algorithmic}\label{IHT_algo2}
\end{algorithm}

\section{Computational Complexity}\label{sec:complexity}
The computational complexity of the reconstruction algorithms compared in this paper was estimated as the number of complex multiplications they need.

The complexity of the proposed Weighted IHT algorithm comes mainly from the calculation of the pseudo-inverse $\left(\mathbf{A}_\mathcal{S}^\mathrm{H}\mathbf{A}_\mathcal{S}\right)^{-1}\mathbf{A}_\mathcal{S}^\mathrm{H}\overline{\mathcal{Y}}$, which is $\mathcal{O}\left(P\left|\mathcal{S}\right|^2\right)$.

The W-SABMP algorithm needs to compute this pseudo-inverse several times, but the efficient implementation proposed in Section IV of~\cite{Masood2012} reduces its complexity to $\mathcal{O}\left(P\left|\mathcal{S}\right|\right)$ by exploiting the previous results. Even then, the calculation needs to be done $N$ times for each possible support size up to $\left|\mathcal{S}\right|$, making the overall complexity of this algorithm $\mathcal{O}\left(NP\left|\mathcal{S}\right|^2\right)$.

Finally, the PANC algorithm performs an FFT and an IFFT of size $N$ in each iteration. We found that two iterations are needed to achieve BER results comparable to the other two algorithms, and therefore PANC has a complexity of 
$\mathcal{O}\left(N\log{N}\right)$.

\section{Simulation Results}\label{sec:results}

The proposed Weighted IHT algorithm was tested and compared with the existing techniques W-SABMP and PANC in terms of both uncoded BER and computational complexity. The performance without recovery and with the oracle-LS method (where the support of the clip signal $\mathbf{c}$ is perfectly known at the receiver and a least-squares solution is applied to estimate its values) are also given for reference. Several experiments were carried out to evaluate the effect of the variation of different scenario parameters on these two performance measures.

All the experiments simulate an OFDM system with $N=512$ subcarriers and 16-QAM modulation. The clipping ratio is defined as:
\begin{equation}
\mathrm{CR}=\frac{\tau}{\sigma_x},
\end{equation}
where $\sigma_x$ is the standard deviation of the input signal to the PA, and $\tau$ is the clipping level. The clipped OFDM signal goes through a 4-tap random complex channel (generated using a Gaussian distribution), and is affected by AWGN. The SNR is defined at the receiver input:
\begin{equation}
\mathrm{SNR}=10\log{\frac{\E{\left[\left\|\mathbf{Hx_p}\right\|_{2}^{2}\right]}}{\sigma_z^2}}
\end{equation}
All the clipping estimation algorithms are applied after channel equalization. The selection of reliable carriers is made according to the geometric reliability measure introduced in~\cite{Al-Safadi2012}, which is computationally faster than the Bayesian approach ($\mathcal{O}(N)$ instead of $\mathcal{O}(NM)$, where $M$ is the QAM modulation order), while getting almost equally good results.

We used the efficient implementation of the W-SABMP algorithm proposed in Section IV of~\cite{Masood2012}. The PANC algorithm~\cite{Gregorio2012} is run for two iterations, as one was not enough to achieve results comparable to the other algorithms. All algorithms estimate the clipping level as the absolute value of the maximum sample of the estimated received signal in the time domain.

\subsection{Experiment I: Performance vs number of reliable carriers}
For the first experiment, the clipping ratio was fixed at $\mathrm{CR}=1.3$, and the $E_b/N_0$ at 15 dB, and the performance of the algorithms was evaluated for varying number of reliable carriers $P$.

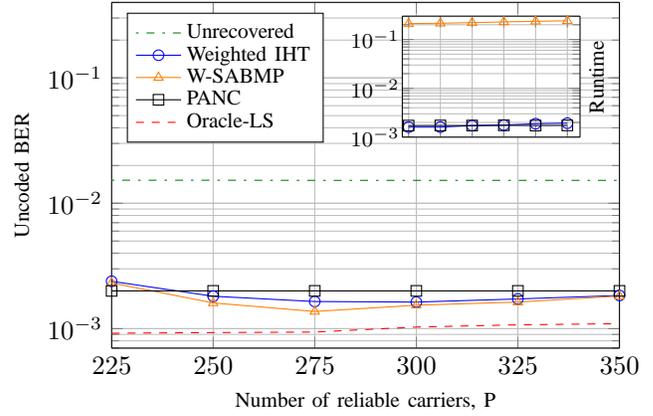
\begin{figure}[!t]\centering
% Minimal pgfplots version: 1.3
%  http://www.mathworks.com/matlabcentral/fileexchange/22022-matlab2tikz
\begin{tikzpicture}

\begin{axis}[%
width=0.95092\figurewidth,
height=\figureheight,
at={(0\figurewidth,0\figureheight)},
name=final_ber_vs_P,
scale only axis,
separate axis lines,
every outer x axis line/.append style={black},
every x tick label/.append style={font=\color{black}},
xmin=225,
xmax=350,
xtick={225, 250, 275, 300, 325, 350},
xlabel={Number of reliable carriers, P},
xmajorgrids,
every outer y axis line/.append style={black},
every y tick label/.append style={font=\color{black}},
ymode=log,
ymin=0.0007,
ymax=0.4,
yminorticks=true,
ylabel={Uncoded BER},
ymajorgrids,
yminorgrids,
legend style={at={(0.03,0.97)},anchor=north west,legend cell align=left,align=left,fill=white, row sep=-0.25em}
]
\addplot [color=black!50!green,dash pattern=on 1pt off 3pt on 3pt off 3pt]
  table[row sep=crcr]{%
225	0.015270144440407\\
250	0.015270144440407\\
275	0.0152304006177326\\
300	0.0152304006177326\\
325	0.0152304006177326\\
350	0.0152304006177326\\
};
\addlegendentry{Unrecovered};

\addplot [color=blue,solid,mark=o,mark options={solid}]
  table[row sep=crcr]{%
225	0.00238746820494186\\
250	0.00181402162063954\\
275	0.00164652979651163\\
300	0.00162949672965117\\
325	0.00172601744186047\\
350	0.00183957122093023\\
};
\addlegendentry{Weighted IHT};

\addplot [color=orange,solid,mark=triangle,mark options={solid}]
  table[row sep=crcr]{%
225	0.00231649709302326\\
250	0.00160394712936047\\
275	0.00137116188226744\\
300	0.001539\\
325	0.00163\\
350	0.00182\\
};
\addlegendentry{W-SABMP};

\addplot [color=black,solid,mark=square,mark options={solid}]
  table[row sep=crcr]{%
225	0.00200138535610465\\
250	0.00200138535610465\\
275	0.00200138535610465\\
300	0.00200138535610465\\
325	0.00200138535610465\\
350	0.00200138535610465\\
};
\addlegendentry{PANC};

\addplot [color=red,dashed]
  table[row sep=crcr]{%
225	0.000919785610465117\\
250	0.000931140988372093\\
275	0.000939657521802326\\
300	0.00103050054505814\\
325	0.0010730832122093\\
350	0.00109579396802326\\
};

\addlegendentry{Oracle-LS};

\end{axis}

\begin{axis}[%
axis background/.style={fill=white},
width=0.95092\subgraphwidth,
height=\subgraphheight,
anchor=outer north east,
at={($(final_ber_vs_P.north east)+(-0.25em,-0.5em)$)},
scale only axis,
separate axis lines,
every outer x axis line/.append style={black},
every x tick label/.append style={font=\color{black}},
xmin=220,
xmax=360,
xtick={225, 250, 275, 300, 325, 350},
xticklabel=\empty,
xmajorgrids,
every outer y axis line/.append style={black},
every y tick label/.append style={xshift=0.25em, anchor=east, font=\color{black} \footnotesize},
ymode=log,
ymin=0.001,
ymax=0.3,
ytick={1e-3, 1e-2, 1e-1},
yminorticks=true,
ylabel right={Runtime},
ymajorgrids,
yminorgrids
]
\addplot [color=blue,solid,mark=o,mark options={solid},forget plot]
  table[row sep=crcr]{%
225	0.00160344267237416\\
250	0.00160672120433308\\
275	0.00172772843349605\\
300	0.0017627766528612\\
325	0.00188720495885086\\
350	0.00193451649137212\\
};
\addplot [color=orange,solid,mark=triangle,mark options={solid},forget plot]
  table[row sep=crcr]{%
225	0.210794343748374\\
250	0.214249796160839\\
275	0.220774947846396\\
300	0.228268611727024\\
325	0.235255911692029\\
350	0.240514013229589\\
};
\addplot [color=black,solid,mark=square,mark options={solid},forget plot]
  table[row sep=crcr]{%
225	0.00172616043995048\\
250	0.00173443249680878\\
275	0.00172940779021957\\
300	0.00175017479564411\\
325	0.00172864161155526\\
350	0.00174131474120623\\
};
\end{axis}

\end{tikzpicture}%
\caption{BER vs P for $\mathrm{CR}=1.3$ and $E_b/N_0=15 \;\mathrm{dB}$.}
\label{fig:final_ber_vs_P}
\end{figure}

Fig. ~\ref{fig:final_ber_vs_P} shows the results. As expected, the execution time of Weighted IHT and W-SABMP increases linearly with $P$. Our proposed technique executes in the order of $N$ times faster than the efficient implementation of W-SABMP, and has a comparable speed to that of PANC. In terms of BER, the sparse recovery algorithms benefit from an increase of the number of reliable subcarriers, up to a point where too many subcarriers with errors are incorrectly selected as reliable and the performance degrades again. The best results seem to be around $P=275$, so we chose this value for the remaining experiments.

\subsection{Experiment II: Performance vs $E_b/N_0$}
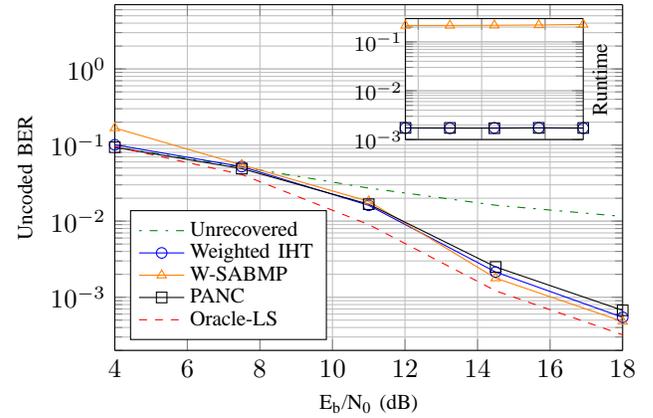
\begin{figure}[!t]\centering
% Minimal pgfplots version: 1.3
%  http://www.mathworks.com/matlabcentral/fileexchange/22022-matlab2tikz
\begin{tikzpicture}

\begin{axis}[%
name=final_ber_vs_snr,
width=0.95092\figurewidth,
height=\figureheight,
at={(0\figurewidth,0\figureheight)},
scale only axis,
separate axis lines,
every outer x axis line/.append style={black},
every x tick label/.append style={font=\color{black}},
xmin=4,
xmax=18,
xlabel={$\text{E}_\text{b}\text{/N}_\text{0}\text{ (dB)}$},
xmajorgrids,
every outer y axis line/.append style={black},
every y tick label/.append style={font=\color{black}},
ymode=log,
ymin=0.0002,
ymax=7,
yminorticks=true,
ylabel={Uncoded BER},
ymajorgrids,
yminorgrids,
legend style={at={(0.03,0.03)},anchor=south west,legend cell align=left,align=left,fill=white, row sep=-0.25em}
]
\addplot [color=black!50!green,dash pattern=on 1pt off 3pt on 3pt off 3pt]
  table[row sep=crcr]{%
4	0.0937137276785715\\
7.5	0.0514285714285714\\
11	0.02716796875\\
14.5	0.0161746651785714\\
18	0.0115234375\\
};
\addlegendentry{Unrecovered};

\addplot [color=blue,solid,mark=o,mark options={solid}]
  table[row sep=crcr]{%
4	0.100717075892857\\
7.5	0.0520954241071429\\
11	0.0161216517857143\\
14.5	0.00214285714285714\\
18	0.000544084821428572\\
};
\addlegendentry{Weighted IHT};

\addplot [color=orange,solid,mark=triangle,mark options={solid}]
  table[row sep=crcr]{%
4	0.167611607142857\\
7.5	0.0548158482142857\\
11	0.0182170758928572\\
14.5	0.00178292410714286\\
18	0.00048\\
};
\addlegendentry{W-SABMP};

\addplot [color=black,solid,mark=square,mark options={solid}]
  table[row sep=crcr]{%
4	0.0937332589285715\\
7.5	0.0489899553571428\\
11	0.0166908482142857\\
14.5	0.0025\\
18	0.000672433035714286\\
};
\addlegendentry{PANC};

\addplot [color=red,dashed]
  table[row sep=crcr]{%
4	0.0956305803571428\\
7.5	0.0410072544642857\\
11	0.00901506696428571\\
14.5	0.00121372767857143\\
18	0.000320870535714286\\
};
\addlegendentry{Oracle-LS};

\end{axis}

\begin{axis}[%
axis background/.style={fill=white},
anchor=outer north east,
at={($(final_ber_vs_snr.north east)+(-0.25em,-0.5em)$)},
width=0.95092\subgraphwidth,
height=\subgraphheight,
scale only axis,
separate axis lines,
every outer x axis line/.append style={black},
every x tick label/.append style={font=\color{black}},
xmin=4,
xmax=18,
xticklabel=\empty,
xmajorgrids,
every outer y axis line/.append style={black},
every y tick label/.append style={xshift=0.25em, anchor=east, font=\color{black}\footnotesize},
ymode=log,
ymin=0.001,
ymax=0.3,
yminorticks=true,
ylabel right={Runtime},
ymajorgrids,
yminorgrids
]
\addplot [color=blue,solid,mark=o,mark options={solid},forget plot]
  table[row sep=crcr]{%
4	0.00172571498723867\\
7.5	0.00170892587453058\\
11	0.00173304713887505\\
14.5	0.00173160387208879\\
18	0.00170862742121367\\
};
\addplot [color=orange,solid,mark=triangle,mark options={solid},forget plot]
  table[row sep=crcr]{%
4	0.215897494560131\\
7.5	0.216416870149465\\
11	0.217932683364372\\
14.5	0.220857463506732\\
18	0.226496431567416\\
};
\addplot [color=black,solid,mark=square,mark options={solid},forget plot]
  table[row sep=crcr]{%
4	0.00173291795758863\\
7.5	0.00173134550951594\\
11	0.00170666742928171\\
14.5	0.00173461513242062\\
18	0.00172850797574171\\
};
\end{axis}

\end{tikzpicture}%
\caption{BER vs $E_b/N_0$ for $\mathrm{CR}=1.3$ and $P=275$.}
\label{fig:final_ber_vs_snr}
\end{figure}

With a clipping ratio of $\mathrm{CR}=1.3$, and $P=275$ reliable carriers, the performance of the three algorithms was measured for different values of $E_b/N_0$, obtaining the plots in Fig.~\ref{fig:final_ber_vs_snr}. The BER results for Weighted IHT are only slightly worse than those for W-SABMP, with a computational complexity two orders of magnitude lower. This confirms that the proposed weighting $\rho_i=\exp{\left\{-\left(\hat{\tau}-\hat{x}_p\left[i\right]\right)\right\}}$ greatly increases the convergence speed of these two algorithms, making the benefits of the other factor $p\left(\mathcal{Y}|\mathcal{S}\right)$ marginal, and in most cases not worth the considerable additional computational cost at which they come.

Our technique also obtains comparable or better BER results than PANC, especially at high SNR, where the reliability measure is more likely to be correct. The speed of both algorithms is almost the same with these parameters, and will be better for Weighted IHT if $N$ is reduced (because $\left|\mathcal{S}\right|$ will decrease proportionally) or if the clipping ratio increases. Even with unfavorable conditions (high $N$ and low clipping ratio), the difference in speed is very small, and tolerable given the fact that Weighted IHT does not need to have the PA model of the transmitter, which is not readily available in the uplink of a mobile communications channel.

\subsection{Experiment III: Performance vs CR}
\begin{figure}[!t]\centering
% Minimal pgfplots version: 1.3
%  http://www.mathworks.com/matlabcentral/fileexchange/22022-matlab2tikz
\begin{tikzpicture}

\begin{axis}[%
name=final_ber_vs_tau,
width=0.95092\figurewidth,
height=\figureheight,
at={(0\figurewidth,0\figureheight)},
scale only axis,
separate axis lines,
every outer x axis line/.append style={black},
every x tick label/.append style={font=\color{black}},
xmin=1.2,
xmax=1.6,
xlabel={$\text{Clipping ratio, CR=}\tau\text{/}\sigma{}_\text{x}$},
xmajorgrids,
every outer y axis line/.append style={black},
every y tick label/.append style={font=\color{black}},
ymode=log,
ymin=0.0001,
ymax=1.75,
yminorticks=true,
ylabel={Uncoded BER},
ymajorgrids,
yminorgrids,
legend style={at={(0.03,0.97)},anchor=north west,legend cell align=left,align=left,fill=white, row sep=-0.25em}
]
\addplot [color=black!50!green,dash pattern=on 1pt off 3pt on 3pt off 3pt]
  table[row sep=crcr]{%
1.2	0.0347711267605634\\
1.3	0.0154805787852113\\
1.4	0.00631327024647887\\
1.5	0.00239670444542254\\
1.6	0.000849334286971831\\
};
\addlegendentry{Unrecovered};

\addplot [color=blue,solid,mark=o,mark options={solid}]
  table[row sep=crcr]{%
1.2	0.0133245763644366\\
1.3	0.0017640019806338\\
1.4	0.000371368838028169\\
1.5	0.000209754621478873\\
1.6	0.000178807218309859\\
};
\addlegendentry{Weighted IHT};

\addplot [color=orange,solid,mark=triangle,mark options={solid}]
  table[row sep=crcr]{%
1.2	0.0169488611355634\\
1.3	0.00147515955105634\\
1.4	0.000261333626760564\\
1.5	0.000223509022887324\\
1.6	0.000196000220070423\\
};
\addlegendentry{W-SABMP};

\addplot [color=black,solid,mark=square,mark options={solid}]
  table[row sep=crcr]{%
1.2	0.0161029654489437\\
1.3	0.00217319542253521\\
1.4	0.000374807438380282\\
1.5	0.000182245818661972\\
1.6	0.000144421214788733\\
};
\addlegendentry{PANC};

\addplot [color=red,dashed]
  table[row sep=crcr]{%
1.2	0.0089266065140845\\
1.3	0.00101094850352113\\
1.4	0.000196000220070423\\
1.5	0.000151298415492958\\
1.6	0.000151298415492958\\
};
\addlegendentry{Oracle-LS};

\end{axis}

\begin{axis}[%
axis background/.style={fill=white},
anchor=outer north east,
at={($(final_ber_vs_tau.north east)+(-0.25em,-0.5em)$)},
width=0.95092\subgraphwidth,
height=\subgraphheight,
scale only axis,
separate axis lines,
every outer x axis line/.append style={black},
every x tick label/.append style={font=\color{black}},
xmin=1.2,
xmax=1.6,
xticklabel=\empty,
xmajorgrids,
every outer y axis line/.append style={black},
every y tick label/.append style={xshift=0.25em, anchor=east, font=\color{black}\footnotesize},
ymode=log,
ymin=0.0005,
ymax=0.4,
yminorticks=true,
ylabel right={Runtime},
ymajorgrids,
yminorgrids
]
\addplot [color=blue,solid,mark=o,mark options={solid},forget plot]
  table[row sep=crcr]{%
1.2	0.00249453964065864\\
1.3	0.00172483744539641\\
1.4	0.00121530636099345\\
1.5	0.000892687234465639\\
1.6	0.000662642090506013\\
};
\addplot [color=orange,solid,mark=triangle,mark options={solid},forget plot]
  table[row sep=crcr]{%
1.2	0.342844831180549\\
1.3	0.22191773005138\\
1.4	0.143227096425188\\
1.5	0.0917961998607337\\
1.6	0.055582493565881\\
};
\addplot [color=black,solid,mark=square,mark options={solid},forget plot]
  table[row sep=crcr]{%
1.2	0.00177296861090739\\
1.3	0.00175501686662148\\
1.4	0.00171883274284122\\
1.5	0.0017694985342824\\
1.6	0.00173952847583187\\
};
\end{axis}

\end{tikzpicture}%
\caption{BER vs CR for $E_b/N_0=15 \;\mathrm{dB}$ and $P=275$.}
\label{fig:final_ber_vs_tau}
\end{figure}
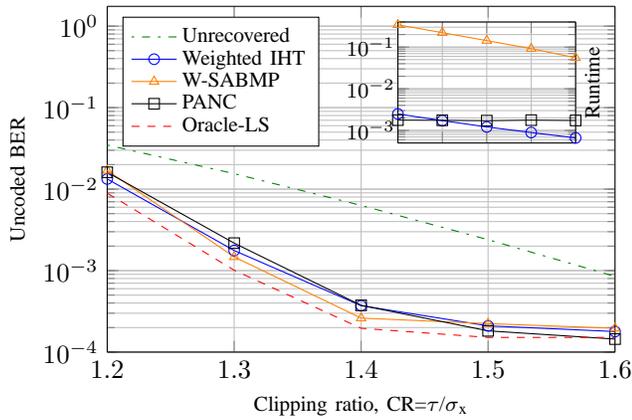

In this experiment, we aimed to check the range of values of the clipping ratio for which our proposed algorithm gives good results. We fixed the $E_b/N_0$ at 15 dB and $P=275$ reliable carriers, and plotted the BER and runtime over CR, as shown in Fig.~\ref{fig:final_ber_vs_tau}. Again, W-SABMP performs only marginally better than Weighted IHT in terms of BER, supporting our idea that the weighting is the main contribution to their convergence rate. The results are also comparable to PANC, slightly outperforming it at severe clipping levels and somewhat worse at higher CR.

As for the execution time, a strong inverse dependence with the clipping ratio can be observed for the sparse recovery algorithms, due to the quadratic relationship between their computational complexity and the size of the support set of the clip signal $\left|\mathcal{S}\right|$, which decreases when CR increases. Weighted IHT is still two orders of magnitude faster than W-SABMP, and from the two graphs we can verify that it is also advantageous with respect to PANC when the clipping ratio is low (no higher than 1.4).

\section{Conclusion}\label{sec:conclusion}

An weighted iterative hard thresholding algorithm for mitigating the nonlinear effects of power amplifiers  is proposed. Compared to the existing PANC technique~\cite{Gregorio2012}, our algorithm obtains similar results, with slightly better BER performance and slower execution for milder clipping, and vice versa for more severe clipping. However, unlike PANC, our technique is able to recover even phase distortion without knowing the PA model, which is not available in the uplink of a mobile communication cell. In this scenario, Weighted IHT with similar performance but fewer assumptions can be advantageous with respect to the existing algorithms.

The proposed technique also leaves some areas open for future work, which are mainly the optimization of the weighting function and of the number of reliable carriers to select depending on the parameters of the scenario.


%\section*{Acknowledgment}



\bibliographystyle{IEEEtran}
\bibliography{Paper}
%\begin{thebibliography}{1}
  %0.5em minus 0.4em\relax Harlow, England: Addison-Wesley, 1999.
%
%\end{thebibliography}

\end{document}